\documentclass[12pt]{article}
\newcommand{\plabel}{\label}

\def\subdef#1{\gdef\globalColor##1{##1}}      
      \subdef{Black}

\newcommand{\epo}{e^+_1}
\newcommand{\emo}{e^-_1}

\newcommand{\Epz}{E^+_0}
\newcommand{\Emz}{E^-_0}
\newcommand{\pz}{\partial_0}

\usepackage{epsfig}
\usepackage{amsmath}
\usepackage{amssymb} 
\usepackage{cite}

\begin{document}

\begin{titlepage}
\renewcommand{\thefootnote}{\fnsymbol{footnote}}

\hfill TUW-98-19 \\
\begin{center}
\vspace{1cm}

\textbf{\Large{Integrating Geometry in General 2D Dilaton 
Gravity with Matter }}
\vspace{1cm}

\textbf{W.\ Kummer$^{1}$\footnotemark[1], H.\ Liebl$^{1}$\footnotemark[2], 
 and D.V.\ Vassilevich$^{2}$\footnotemark[3]}


{$^1$Institut f\"ur
    Theoretische Physik, Technische Universit\"at Wien \\ Wiedner
    Hauptstr.\  8--10, A-1040 Wien \\ Austria}

{$^2$Institut f\"ur
    Theoretische Physik, Universit\"at Leipzig \\
    Augustusplatz 10, D-04109 Leipzig \\ Germany}
\vspace{1cm}

\footnotetext[1]{e-mail: {\tt wkummer@tph.tuwien.ac.at}}
\footnotetext[2]{e-mail: {\tt liebl@tph16.tuwien.ac.at}}
\footnotetext[3]{e-mail: {\tt Dmitri.Vassilevich@itp.uni-leipzig.de}
\  \  On leave from Dept. of Theoretical Physics,  
St.\ Petersburg University,  
198904 St.\ Petersburg, Russia}

\end{center}
\vfill

\begin{abstract}

General 2d dilaton theories, containing spherically symmetric 
gravity and hence the Schwarzschild black hole as a special case, 
are quantized by an exact path integral of their geometric 
(Cartan-) variables. Matter, represented by minimally coupled 
massless scalar fields is treated in terms of a systematic 
perturbation theory. The crucial prerequisite for our approach is 
the use of a temporal gauge  for the spin connection and for 
light cone components of the zweibeine which amounts to an 
Eddington Finkelstein gauge for the metric. 
We derive the generating functional in its most general form which
allows a perturbation theory in the scalar fields. The relation of the
zero order functional to the classical solution is established. As an
example we derive the effective (gravitationally) induced 4-vertex for
scalar fields.
\end{abstract}

\end{titlepage}

\section{Introduction}

Already for a long time the unsolved problem of quantizing 
gravity has been thought to allow a point of attack in the realm 
of two-dimensional theories  of gravity. After all, the 
Schwarzschild solution for the black hole (BH) can be treated by 
reducing 4d gravity in terms of spherical coordinates 
\cite{tho84}. However, even in that reduced phase space, 
interactions with scalar matter --- moreover simplified to minimal 
coupling --- seemed to be not managable as a quantum theory. If 
studies desired to reach beyond semiclassical approaches 
eventually they had to rely again on computations in a given 
curved background. This is the case also for several most
recent papers dealing with this subject \cite{1a}.

The two-dimensional dilaton BH \cite{man91} (DBH) for the first 
time provided hope that some break-through could be 
achieved: The classical solvability of a theory with a global 
structure coinciding with  the Penrose diagram for the 
Schwarzschild BH seemed to offer the possibility that by 
integrating out the geometric variables (metric) exactly, matter 
could be treated as a perturbation  in a systematic manner. This 
hope was supported by the fact that the DBH model had been 
derived from string theory and thus powerful methods of conformal 
field theory were expected to be applicable in the quantization 
program \cite{banks}. The use of the conformal gauge in all this 
work therefore seemed to be most convenient. Nevertheless, also 
within the 2d DBH the program of a perturbative treatment of 
matter in a quantum exact (nonperturbative) integration of the 
geometry could not be achieved. One loop matter, represented by a 
Polyakov term was plugged back into the classical equations of 
motion, again leading to a basically semiclassical --- albeit 
improved --- treatment of quantum effects around a given more 
dynamical background representing a model of the BH. Also right 
from the beginning of those studies the authors working in this 
field were well aware 
of a basic shortcoming of the DBH with respect to Hawking 
radiation: In that model Hawking temperature and Hawking flux are 
not determined by the black hole's mass, but only depend on a 
cosmological constant. As noted later by M.O. Katanaev and two of 
the present authors \cite{kat97} another serious deviation of 
the Schwarzschild BH affecting possible applications
to final black hole evaporation at times of order inverse Planck mass
and thus also
to the 
information paradox, stems from the null-completeness of the 
(non-null incomplete) singularity. 

As a consequence of the flurry of interest generated by the DBH 
of \cite{man91} and in continuation of even earlier attempts, 
also generalized dilaton theories \cite{lou94a}

\begin{equation}
 \plabel{lbegin}
{\cal{L}}_{(2)}=\sqrt{-g}
\left(-X\frac{R}{2}-V(X) + \frac{U(X)}2\, (\nabla X)^2\, \right) 
\end{equation}
with or without (minimal) coupling to matter \cite{lou94a}

\begin{equation}
\plabel{lmatter}
{\cal{L}}^{(m)} = \frac12 \sqrt{-g}g^{\mu \nu}\partial_{\mu}S \partial_{\nu}S 
\end{equation}
received considerable attention. In (\ref{lbegin}) and (\ref{lmatter}) 
$g_{\mu\nu}$  is the 2d metric, 
 $R$ the Ricci scalar. $U$ and $V$ are general functions of the dilaton 
 field $X$. Spherically reduced gravity (SRG) is the special case 
$U_{SRG}=-(2X)^{-1}$, $V_{SRG}=-2$, 
the DBH follows with $U_{DBH}=-(X)^{-1}$, $V_{DBH}=2\lambda^2 X$.
Some aspects of classical and (in particular cases) quantum
integrability of such models were considered recently in
\cite{dA}.

Another development started with the inclusion of nonvanishing 
torsion in a twodimensional model of gravity \cite{kat86}. In 
connection with that model the advantages of a temporal gauge 
($\omega_0 = 0$) for the components of the spin connection  
$\omega^a_{\mu b}=\omega_{\mu} {\varepsilon^a}_b$ and for the light 
cone components of the zweibein $e^a_{\mu}$
($e_0^+=e_0^- -1=0$) were first realized 
in connection with its classical solution \cite{kumschPR}
and for the path integral in the quantum case \cite{kum92}.
The Dirac quantization could also be carried through 
\cite{sch94} for that model. 
Because this gauge leads to an Eddington Finkelstein form of the 2d
metric, it will be refered to as EF-gauge below.

Closely related to the intriguing properties of that gauge is the fact
that 
actually {\it all} matterless 
2d-models of gravity with arbitrary powers of torsion and 
curvature can be summarized in an action  \cite{str94}

\begin{eqnarray}
  \plabel{lfirst}
  {\cal{L}}_{(2)}&=&\int_{M_2}
\left[ X^+De^-+ X^-De^++Xd\omega +\epsilon {\mathcal{V}}(X^+X^-,X)
  \right] \quad , \\
\plabel{lfirstb}
{\mathcal{V}}&=&X^+X^-U(X) + V(X) \quad 
\end{eqnarray}
where $De^a=de^a+(\omega \wedge e)^a$ is the torsion two form,
the scalar curvature
$R$ is related to the spin connection $\omega$ by $-\frac{R}{2}=* d\omega$
and $\epsilon$ denotes the volume two form
$\epsilon=\frac{1}{2}\varepsilon_{ab}e^a \wedge e^b=d^2x   \det
e^a_{\mu}=d^2x\,e$.  Our conventions are determined by $\eta=diag(1,-1)$
and $\varepsilon ^{ab}$ by $\varepsilon ^{01}=-\varepsilon ^{10}=1$. We
also have to stress that even with Greek indices, $\varepsilon^{\mu \nu}$
 is always understood to be the antisymmetric Levi-Civit\'a 
symbol and never 
the corresponding tensor.
In \cite{kum97}
we have shown that (\ref{lfirst}) is quantum equivalent to the generalized 
dilaton theory (\ref{lbegin}) with $U$ and $V$ representing the same 
functions in both actions. 
This represented the generalization of the classical equivalence 
first established
for the model of \cite{kat86} with a dilaton theory in
\cite{kat96}. 
It should be noted that a removal of the kinetic term 
in (\ref{lbegin}) by a conformal transformation was 
 avoided in that step,  
because this 
would imply drastic changes of global properties already at the 
classic level \cite{kat96}. The essential differences between 
quantum theories related by conformal transformations has been 
pointed out repeatedly, e.g.\ in ref. \cite{kuc97}. It is 
especially relevant for the DBH which otherwise could be 
reduced formally to a theory in a flat background. This is the 
origin of its classical solvability, however did not turn out 
to be helpful to 
obtain a quantum solution in the end. 

In previous work the present authors \cite{nonpert} have shown 
that the strategy of an exact path integral for the geometric 
variables with matter as a perturbation can be pursued to obtain 
2-loop corrections in the scalar field for the class of theories 
(\ref{lfirst}) with $U(X) = 0$. This exercise showed how our approach 
can be used to determine loop corrections for the Polyakov term. 
But the restriction to $U(X) = 0$ in that work eliminated 
dilaton theories \cite{lou94a} and especially also those 
``physical'' ones where vanishing of the absolutely conserved
quantity $\cal C$ \cite{kat86,kumschPR,kum92,str94,kat96,kumwidPR}
implies (classically) a flat background.  DBH and Schwarzschild
BH (SRG) belong precisely to this latter class.

In the present paper we report the successful extension of our
formalism to an {\it arbitrary} theory (\ref{lfirst}) with kinetic
term for the dilaton field ($U \neq 0$). 
To the best of our knowledge therefore here for the first
time a systematic quantization of gravity is available
at least in $d=2$, including even the S-wave part of $d=4$
General Relativity, i.e. the quantum theory of the Schwarzschild 
BH. 
We stress that we achieve this goal without any
additional assumption on the quantum behavior of the BH 
\cite{BeMu}, remaining strictly within ``orthodox'' quantum
field theory.

An important part of our paper also deals with aspects which 
are related to the conserved quantity $\cal C$ referred to 
above. It is found to be intimately connected with  
'homogeneous' solutions of first order differential equations 
which we had not considered in \cite{nonpert} 
 but which represent the classical background to zero order in
the scalar loops. An arbitrariness in the generating functional
leads us to introduce a contribution which generalizes the
corresponding Lagrangian in the second reference of \cite{kum92}
for the special model \cite{kat86}. It turns out to be the only
surviving term if the sources, introduced for the momenta, are
set to zero.

In Section 2 we perform the path integral  to the point 
when the matter fields S are to be integrated. As expected from 
our previous work the 'geometric' part of the path integral can 
be done exactly --- and is trivial in the sense that only 
classical effects remain, if --- as a consequence of our specific 
regularization --- global quantum fluctuations are 
suppressed.  Quantum interactions are then induced by the 
matter fields only.

Contact to the classical conservation law and to the 
classical (EF) solution of the zweibeine and spin connection is 
made in Section 3. In the
(Gaussian) integral of the scalars which is the subject of Section 4
we have to confront the problem of
the (generalized) Polyakov term in the EF gauge. In that gauge a
technical problem arises for  the determinant defining the measure
and for the (effective) d'Alembertian. This we solve in two ways.
We either introduce a
further auxiliary field, acting as a token contribution for that 
component of the metric which in the EF gauge cancels anyhow in the kinetic
term of the $S$-fields. The other approach uses path integrals of
ad hoc ghost fields. 
In the present paper we try to attain a sufficient level of rigor. 
This is why we pay  attention to the asymptotics 
and the path integral measure. We also give two different forms
of generating functional for the effective matter theory because
different forms may appear to be more convenient for different 
applications.

In Section 5 we derive the four vertex of the scalar fields,
induced by gravity.

Our results are summarized in Section 6, together with an outlook
on the wide range of possible applications of our approach. Also
several open (but in principal solvable) problems are listed.

In Appendix A the absence of anomalous
contributions to the measure resulting from the geometric variables 
is demonstrated in the background field formalism.
Appendix B discusses some aspects of the 
UV and IR regularization. 

\section{Dilaton Path Integral with Scalar Matter}

Expressing (\ref{lfirst}) in terms of the components $e_\mu^a, \omega_\mu$ 
and rewriting the matter Lagrangian (\ref{lmatter}) as 
\begin{equation}
\plabel{L-matter}
{\cal{L}}^{(m)}  
=- \frac12 \frac{\varepsilon^{\alpha \mu} \varepsilon^{\beta \nu}}{e}
\eta_{ab}e^a_{\mu}e^b_{\nu}\partial_{\alpha}S \partial_{\beta} S 
\end{equation}
the extended Hamiltonian in the sense of Batalin and 
Vilkovisky \cite{brs} can be constructed following the line described in our 
previous work \cite{kum97} for the matterless 
case. In the EF gauge 
\begin{equation}
\plabel{gauge-fix}
e_0^+= \omega_0=0  ,\quad      e_0^- =1
\end{equation}
after integrating out the two types of ghosts, the gauge-fixed ``coordinates''
(\ref{gauge-fix}) and their respective canonical momenta (primary 
constraints), the generating functional for Green functions reads

\begin{equation}
\plabel{W-begin}
W=\int(dS)(dP)(d^3q)(d^3p) \sqrt{\det \left( \frac{q_3}{q_2}  \right)}
\det F \exp i\int 
\left( \frac {{\cal{L}}_{(1)}^{eff}}{\hbar} +
  {\cal{L}}_{(s)}\right)d^2x \quad ,
\end{equation}
where some explanation of the notation is needed.
The first point is that 
to simplify (\ref{W-begin}) and our formulas below we use 
the shorthand notation for ``coordinates'', ``momenta'' and sources 

\begin{eqnarray}
q_i          &=& (\omega_1, e_1^-, e_1^+) \quad , \nonumber \\
{\bar{q}}_i  &=& (\omega_0, e_0^-, e_0^+) \quad , \nonumber \\
p_i          &=& (X, X^+, X^-) \quad , \plabel{abbrev}\\
j_i          &=& (j, j^+, j^-) \quad , \nonumber \\
J_i          &=& (J, J^-, J^+) \quad . \nonumber
\end{eqnarray}
In the EF gauge 
(\ref{gauge-fix}) $q_3=e_1^+=\det e =\sqrt{-g}$.
Together with the only other nonvanishing component of the zweibein
$q_2=e_1^-$  the product $2q_2q_3$ represents the Killing norm
of $g_{\mu \nu}$ in that gauge. Furthermore, 
in (\ref{W-begin}) the introduction of  
$\left[\det \left(q_3/q_2\right)\right]^{1/2}$
is a consequence of  the 
required covariance of the measure for the final S-integration 
\cite{fuj88}. 

The second determinant 
\begin{equation}
\plabel{F}
\det F = \det [\partial_0 + p_2 U(p_1)]
\end{equation} 
is a remnant of the preceding functional integrations 
\cite{nonpert} in extended phase space,
 which include the ones with respect to $\bar{q}_i$ in
(\ref{abbrev}), fixed by the gauge-fermion \cite{brs} according to 
(\ref{gauge-fix}). Another careful treatment of the Faddeev--Popov
determinant in the background field approach can be found in 
Appendix A, where the absence of gauge anomalies for the measure 
is demonstrated .
Finally in (cf. (\ref{lfirstb}) for the definition of $\mathcal{V}$)
\begin{equation}
\plabel{L-eff}
{\cal{L}}_{(1)}^{\mbox{\scriptsize eff}}= \dot{q_i}p_i + \dot{S}P + q_1 p_2 
-q_3 \mathcal{V}  
+ \frac{1}{4q_2}\left( P- \partial_1 S\right)^2
\end{equation}
the last three terms are the ones remaining from (minus) the
extended Hamiltonian, and $P=\partial {\cal{L}}^{(m)}/ \partial \dot{S}$. 
The source term reads 
\begin{equation}
\plabel{L-source}
{\cal{L}}_{(s)}= j_i q_i +J_i p_i +SQ \quad .
\end{equation}
We also retain $\hbar \neq 1$ in order to keep track of loop 
orders in a simple manner.

After performing the Gaussian integration with respect to the momenta $P$ 
one arrives at

\begin{equation}
\plabel{W-next}
W=\int(dS)(d^3q)(d^3p) \sqrt{\det q_3 }
\det F \exp i\int 
\left( \frac {{\cal{L}}_{(2)}^{\mbox{\scriptsize{eff}}}}{\hbar} +
  {\cal{L}}_{(s)}\right)d^2x 
\end{equation}
where  as compared to (\ref{L-eff}) the new effective Lagrangian
is
\begin{equation}
\plabel{L-eff-2}
{\cal{L}}_{(2)}^{\mbox{\scriptsize{eff}}} = -q_i\dot{p_i}+q_1 p_2 
-q_3 \mathcal{V}  - q_2 (\partial_0 S)^2+
(\partial_0 S)(\partial_1 S) \ .
\end{equation}
The cancellation of the determinant of $q_2$ in the measure
should be noted. 
It is well known that the correct diffeomorphism invariant
measure for a scalar field $S$ on a curved background $e_\mu^a$
is $d((-g)^{1/4} S) = (d \sqrt{e} S)$. 
Note, that $e=e^+_1=q_3$ in the EF gauge. 
Therefore (\ref{W-next}) indeed contains the correct measure. 
We want to be
able to use the same order of simple integrations  as in the
matterless case (first $\int (d^3q)$, then $\int (d^3p)$). The factor 
$\sqrt{\det q_3}$ prohibits this for an immediate $q_3$ integration.

One way (approach a)) to solve this problem consists by introducing a 
new field $f$ and by representing the path integral (\ref{W-next}) as
\begin{equation}
W=\int (df) \delta(f-\frac1i \frac{\delta}{\delta j_3}) \widetilde W 
\label{tilS}
\end{equation}
where in
\begin{equation}
\widetilde W= \int (d\tilde S)(d^3q)(d^3p)\det F \exp i 
\int \left[ \left(\frac{{\cal{L}}_{(2)}^{\mbox{\scriptsize{eff}}}}{\hbar}+
{\cal{L}}_{(s)}\right)d^2x \right]_{S=f^{-\frac12}\, \tilde{S}}
\plabel{tilS-2}
\end{equation}
the $\sqrt{\det f}$ has been absorbed in a new variable 
$\tilde S$ for the scalars.
The representation (\ref{tilS})
allows us to integrate the geometric variables 
as in the matterless case since the action
remains linear in the $q_i$. An important feature of (\ref{tilS-2}) is
that $(d\tilde S)$ is just the standard Gaussian measure independent
of $q_i$. 
Integrating out the $S$-field in section 4, we therefore 
will be able to use the definition of the Polyakov
action
\begin{eqnarray}
&&\int (d\tilde S)\exp \left( i\int d^2x\,
\frac {\tilde S}{g^{1/4}} \partial_\mu \sqrt{g} g^{\mu\nu}
\partial_\nu \frac {\tilde S}{g^{1/4}} \right) =
{\det}^{-\frac 12} \left( \frac {1}{g^{1/4}} 
\partial_\mu \sqrt{g} g^{\mu\nu}
\partial_\nu \frac {1}{g^{1/4}} \right)
\nonumber \\
&&\qquad \qquad =\exp i \int d^2\, x ({\cal L}_P(g_{\mu\nu}))
\label{defP}
\end{eqnarray}
where $g_{\mu\nu}$ are {\it arbitrary} functions of sources and
fields, which should only be independent of $S$.

Another way (approach b)) to achieve the same technical advantage
starts from the path integral identity 
\begin{equation}
\plabel{b17}
\sqrt{det q_3}=\int (d\phi )(dc)(d\bar c) \exp \left(
i\int q_3(\phi^2 +\bar cc) \right) \ ,
\end{equation}
where $\phi$ is a scalar, $c$ and $\bar c$ represent anticommuting
Grassmann fields. Then instead of (\ref{tilS})
\begin{equation}
W=\int (d\phi )(dc)(d\bar c) \widetilde{\widetilde W}
\plabel{b18}
\end{equation}
where
\begin{equation}
\widetilde{\widetilde W}=\int (dS)(d^3q)(d^3p) \det F
\exp i \int d^2x \left( 
\frac{\widetilde{\widetilde {\cal L}}^{eff}_{(2)}}\hbar 
+{\cal L}_{(s)} \right)
\plabel{b19}
\end{equation}
and
\begin{equation}
\widetilde{\widetilde {\cal L}}^{eff}_{(2)}=
{\cal L}^{eff}_{(2)} + \hbar q_3 \tilde{\tilde l}\ ,\qquad
\tilde{\tilde l}=\phi^2 +\bar cc \ .
\plabel{b20}
\end{equation}

The main advantage of the EF gauge is the fact 
that 
due to the linearity in $q_i$ in both approaches
the $(d^3q)$ integration in 
$\widetilde W$ or $\widetilde{\widetilde W}$
can be done first -- as in the case without matter fields --
leading to three $\delta$-functions: 
\begin{eqnarray}
\plabel{A1}
&{}&\delta\left(-\nabla_0 \left( p_1 -\hat{B_1} \right)\right) \\
\plabel{A2}
&{}&\delta\left(-\nabla_0 \left( p_2 -\hat{B_2} \right)\right) \\
\plabel{A3}
&{}&\delta\left(-F \left( p_3 -\hat{B_3} \right) \right)    \quad .
\end{eqnarray}
where $F$ is the differential operator in (\ref{W-next}). 
$\hat{B}_i$ will be given below. 
The symbol
$\nabla_0 =\partial_0 -i(\mu-i\varepsilon)=\partial_0 -i\tilde{\mu}$ 
describes 
an IR regularized 'derivative', related to Green 
functions $\nabla^{-1}_{0xx'}$, 
$\nabla^{-2}_{0xx'}$
with proper asymptotics (cf. Appendix B). Below we will 
be careful to mark the difference with respect to 
$\tilde{\nabla}_0=\partial_0 + i\tilde{\mu}$
which appears e.g. in partial integrations like 
$\int A (\nabla_0 B) =-\int (\tilde{\nabla}_0 A)B $ \cite{nonpert}. 
Beside the IR and UV asymptotics for a proper definition of
consistent Green functions also the allowed range of the variable
$x^0$ is crucial. For example, a singularity at $x^0=0$ -- as in
the case of the Schwarzschild black hole for our choice of 
coordinates -- require appropriate
boundary conditions for the half line $x^0\ge 0$. This point
will be discussed in more detail in future work. Here we mainly
concentrate on the general formalism.

Using these three $\delta$-functions the remaining integrations over
$(d^3p)$ yield directly  
\begin{equation}
\plabel{pi}
p_i=\hat{B}_i \quad .
\end{equation}
This simply means that in the phase-space (path-) integral
(\ref{W-begin}) only 
classical paths contribute to the $p$-s, fixing them by (\ref{pi}) 
in terms of the sources 
(and some homogeneous solutions, representing a classical
background, as will be demonstrated below).

Solving (\ref{A1}) and (\ref{A2}) we observe that 
the terms $\hat{B}_1$ and $\hat{B}_2$ 
allow for homogeneous solutions 
($\nabla_0 \bar{p}_1=\nabla_0 \bar{p}_2 = 0$) and 
the $S$-dependent parts can be separated easily:
\begin{eqnarray}
\hat{B}_1 &=& \underbrace{\bar{p}_1 + \nabla_0^{-1} \bar{p}_2 + 
\hbar (\nabla_0^{-1}j_1 + \nabla_0^{-2}j_2) } 
 -\nabla^{-2}_0 (\partial_0 S)^2 
\plabel{B12} \\
\nonumber {}&{}& \qquad \qquad \qquad:=B_1 \\
\hat{B}_2 &=& \underbrace{\bar{p}_2 + \hbar \nabla_0^{-1} j_2 }  
-\nabla^{-1}_0 (\partial_0 S)^2 \quad \plabel{B22} \\
\nonumber {}&{}& \qquad :=B_2
\end{eqnarray}
For the computation of $\hat B_3$ it is convenient to use an exponential 
form of the 
differential operator $\hat{F}=F(\hat{B}_1, \hat{B}_2)$ defined in (\ref{F})
\begin{eqnarray}
\hat{F} &=& e^{-\hat{T}}\nabla_0 e^{\hat{T}}  \plabel{Fhat}    \\
\hat{T} &=& \nabla^{-1}_0 (\hat{U} \hat{B}_2) \plabel{That}    \\
\hat{U} &=& U(\hat{B}_1)                      \plabel{Uhat} \quad .
\end{eqnarray}
Including the homogeneous solution from 
the operator $\hat F$  
we deduce $(\nabla_0 \bar{p}_3 = 0)$ for approach a) 
\begin{eqnarray}
\plabel{B32}
\hat{B}_3 &=& e^{-\hat{T}} 
\left[ \nabla_0^{-1}e^{\hat{T}}(\hbar j_3 -V(\hat{B}_1))+\bar{p}_3
\right] \\
{}        &=& \underbrace{
e^{-T}\left[ \nabla_0^{-1}e^{T}(\hbar j_3 -V(B_1))+\bar{p}_3 \right]}   
+{\mbox{terms }}{\cal{O}}(S^2) \ . \\
{}        &{}& \qquad \qquad \qquad :=B_3 \nonumber
\end{eqnarray}
In approach b) only (\ref{B32}) must be replaced by the
expression
\begin{equation}
\widetilde{\widetilde B}_3=e^{-\hat{T}} 
\left[ \nabla_0^{-1}e^{\hat{T}}(\hbar j_3 -V(\hat{B}_1)
+ \hbar\,\tilde{\tilde l})+\bar{p}_3
\right]=\hat B_3 +e^{-\hat{T}}
\nabla_0^{-1}e^{\hat{T}}\hbar\,\tilde{\tilde l}
\plabel{B32-b}
\end{equation} 
As we shall demonstrate in Section \ref{class} the homogeneous
solutions $\bar{p}_i$ will provide the essential ingredient to recover
the classical solutions of the theory.
It should be emphasized 
that by performing the $p_3$ intepration the term
$\det F$ in the path integral measure has been cancelled.
Thus no Faddeev-Popov type 
determinant appears finally in our gauge -- not too surprisingly in view 
of its 'axial' character (for more details on the
measure we refer to Appendix A). 

We now argue that (for approach a)) in 
\begin{eqnarray}
\plabel{Wtilde}
\widetilde{W} &=&\int (dS) \exp i/\hbar \int
{\cal{L}}_{(3)}^{\mbox{\scriptsize{eff}}} \\
\plabel{Wtilde-b}
{\cal{L}}_{(3)}^{\mbox{\scriptsize{eff}}} &=& \hbar J_i \hat{B}_i + \hbar SQ + 
(\partial_0 S)(\partial_1 S) + \hat{\cal{L}}^{HK}
\end{eqnarray}
we have to add another term
\begin{equation}
\plabel{L-HK}
\hat{\cal{L}}^{HK}=\tilde{g} e^{\hat{T}} (\hbar j_3 - \hat{V})
\end{equation}
to the effective Lagrangian ${\cal L}_{(2)}^{eff}$
 which survives the limit of vanishing 
sources $J_i$ of the momenta. It is the result of an 
ambiguity in the treatment of the three terms proportional to $J_i$: 
For $J_1$ and $J_2$ the (nonlocal) $\nabla_0^{-1}$-factors in 
(\ref{B12},\ref{B22}) 
(before adding the homogeneous solution) could also have been attached to 
$J_{1,2}$. 

E.g. for the simplest expression of this type in $J_2B_2$ 
\begin{equation}
\plabel{simplest}
\int_{x'}\int_{x}J_{2x}\nabla^{-1}_{xx'}j_{2x'}=
-\int_{x'}\int_{x}\tilde{\nabla}^{-1}_{xx'}J_{2x'}j_{2x}
\end{equation}
also the r.h.s. of this equation could be taken to represent the
``correct'' one. But then  a homogeneous
solution may be added: 
$\nabla^{-1}J_{2} \to \nabla^{-1}J_{2} -\tilde{g}_2$
with $\tilde{\nabla}_0 \tilde{g_2}=0$.
Proceeding in a similar way for $J_1B_1$
  produces additional contributions $(\widetilde \nabla 
\tilde{g}_1 = 0)$ as 
\begin{equation}
\plabel{J-replace}
J_1 \hat{B}_1 + J_2 \hat{B}_2 \to J_1 \hat{B}_1 + J_2 \hat{B}_2 
+\tilde{g}_1 (j_1 +\nabla^{-1}_0 j_2) + \tilde{g}_2 j_2 \quad .
\end{equation}
These terms, however, must be irrelevant because they would only contribute 
to couplings of the sources of $e_1^-$ and $\omega_1$ to some 
'external fields' $\tilde{g}_{1,2}$. 
The situation is different for a similar contribution from $J_3\hat B_3$. 
There after an analogous reordering 
$\int J_3 \hat{B}_3 =-\int \left(\widetilde{\nabla}^{-1}_0 J_3
  e^{-\hat{T}} \right) e^{\hat{T}}(\hbar j_3 -\hat{V})$
a homogeneous solution $(-\tilde{g}_3 /\hbar)$ is added which
leads to the additional expression (\ref{L-HK}). 
There are several further reasons why $\hat{\cal L}^{HK}$
must be present: 1) 
It is impossible that all the dynamics disappear together with the 
sources of the momenta; 
indeed the classical equations of motion for the $q_i$ (cf. 
(\ref{xvariation-1}) to (\ref{xvariation-3}) below) acquire their 
``quantum'' counterpart from ${\cal{L}}^{HK}$ alone!
 2) For the special case of the 
Katanaev-Volovich model it was shown, keeping there 
$J_i = 0$ from the start and 
following the traditional sequence of integrations
({\it{first}} a Gaussian 
integral of ($d^3p$) and 
{\it{then}} another Gaussian integral for the $\int d^3 q$), that precisely an 
effective action of type (\ref{L-HK}) is produced (cf. the second
ref.\ \cite{kum92}). 3) In the matterless classical case on the mass 
shell by the simple method
explained in section \ref{sec3-1}  $(\hbar j_3 - V)e^{T}$
 may be expressed as the derivative of
the conserved quantity $\cal{C}$. This is reminiscent of $\cal{C}$ 
appearing at the boundary for
 a quasilocal energy definition \cite{kumwidPR}.

For approach b) the same arguments are valid. Here the additional term
$\tilde{\tilde l}$ which enters the total Lagrangian like an
addition to $j_3$  modifies (\ref{L-HK}) to
\begin{equation}
\plabel{L-HK-b}
\widetilde{\widetilde L}^{HK}=\tilde{g} e^{\hat{T}} (\hbar j_3
+ \hbar\,\tilde{\tilde l} - \hat{V})
\end{equation}

\section{Classical and Quantum Equations of Motion, Conservation 
Law} \plabel{class}
As seen already in the previous section, after integrating
the geometric variables the path integral
contains many terms with inverse derivatives $\nabla_0^{-1}$.
These inverse derivatives (regularized $x^0$-integrals)
allow several integration constants, or homogeneous modes.
In this section we relate the homogeneous modes to each other
and to the conserved quantity ${\cal C}$ by means of equations
of motion and Slavnov--Taylor like identities.

Since a nontrivial quantum theory only emerges through the interaction with 
matter, we have to show how 
classical dynamics is reproduced in the path 
integral formalism if  the matter terms are disregarded to zero loop order,
 the order considered in this section.

\subsection{Classical Equations of Motion}
\plabel{sec3-1}

Variation of $\delta \omega_1$, $\delta e_1^{\pm}$ (or $\delta q_i$) 
in (\ref{lfirst}), and  
fixing the gauge according to (\ref{gauge-fix}) yields 
the classical equations of motion for the $p_i$
\begin{eqnarray}
\plabel{1-variation-1}
\nabla_0 p_1 -p_2 &=&0 \\
\plabel{1-variation-2}
\nabla_0 p_2 &=& 0 \\
\plabel{1-variation-3}
\left(\nabla_0 + p_2 U(p_1) \right)p_3 + V(p_1) &=&0 \quad .
\end{eqnarray}
The analogous variation of $\delta \omega_0$, $\delta e_0^{\pm}$  
in (\ref{lfirst}) results in 
\begin{eqnarray}
\plabel{0-variation-1}
\nabla_1 p_1 + p_3 q_3 -p_2 q_2    &=& 0 \\
\plabel{0-variation-2}
\nabla_1 p_2 + p_2 q_1 -(V(p_1)+Up_2p_3)q_3  &=& 0 \\
\plabel{0-variation-3}
(\nabla_1 + p_2 U)p_3 - q_1 p_3 + V q_2 &=& 0   \quad ,
\end{eqnarray}
and finally variation of $\delta X$, $\delta X^{\pm}$ (or $\delta p_i$)  
provides a third triple of equations of motion: 
\begin{eqnarray}
\plabel{xvariation-1}
\widetilde{\nabla}_0 q_1 -p_2p_3 q_3 \frac{\partial U}{\partial p_1}
-q_3 \frac{\partial V}{\partial p_1} &=& 0 \\
\plabel{xvariation-2}
\widetilde{\nabla}_0 q_2 + q_1 -q_3 p_3 U &=& 0 \\
\plabel{xvariation-3}
(\widetilde{\nabla}_0 - U p_2 ) q_3 &=& 0 \quad .
\end{eqnarray}

As can be seen we distinguished the regularized $\nabla_\mu$ from 
$\tilde\nabla_\mu$ appearing after partial integration. 
The peculiar property of all covariant 2d models to 
provide one set of equations involving the momenta alone \cite{gros92} 
is clear from (\ref{1-variation-1}-\ref{1-variation-3}).

As expected, the solution of eqs. 
(\ref{1-variation-1}-\ref{1-variation-3}) coincides with the quantities 
$B_i^{(0)}=B_i(j=0)$ as defined in (\ref{B12},\ref{B22}) and (\ref{B32})
for $j_i=0$ and $S=0$:
\begin{eqnarray}
\plabel{B1-0}
p_1&=&B_1^{(0)}=\bar{p}_1 + \nabla_0^{-1}\bar{p}_2 \\
\plabel{B2-0}
p_2&=&B_2^{(0)}=\bar{p}_2 \\
\plabel{B3-0}
p_3&=&B_3^{(0)}=e^{-T^{(0)}}
\left( \bar{p}_3-\nabla_0^{-1}e^{T^{(0)}}V^{(0)} \right) \\
\plabel{T-0}
T^{(0)}&=&\nabla_0^{-1} \left(B_2^{(0)} U^{(0)} \right) \\
V^{(0)}&=&V(B_1^{(0)}) \\
U^{(0)}&=&U(B_1^{(0)}) \quad .
\end{eqnarray}

The zero component of the absolute conservation law  
\cite{gros92,sch94,str94,kum97,kumschPR,kumwidPR,strklo-CQG,KumTie}
is obtained by linear combination of eqs.\  
(\ref{1-variation-1}-\ref{1-variation-3})
\begin{equation}
\plabel{delta-C}
\nabla_0 (p_2 p_3) + p_2^2 p_3 U + V(\nabla_0 p_1) = 0
\end{equation}
and thus 
\begin{equation}
\plabel{delta-C-b}
\nabla_0 {\cal{C}} = 
\nabla_0 \left[e^{Q(p_1)}p_2 p_3 + w(p_1) \right] = 0
\end{equation}
follows with 
\begin{eqnarray}
\plabel{Q-def}
Q(x) &=& \int_{y_0}^x U(y)dy \\
\plabel{w-def}
w &=& \int_{x_0}^x V(y)e^{Q(y)}dy \quad .
\end{eqnarray}

In an analogous way one derives from the classical eqs.
(\ref{0-variation-1}-\ref{0-variation-3})
$\nabla_1 {\cal C} = 
0$  so that
\begin{equation}
\plabel{C-const}
{\cal{C}}={\cal{C}}_0 = \mbox{const} \quad .
\end{equation}
Therefore $p_2p_3$ and $p_1$ always depend on each other. The 
constant ${\cal  C}_0$ alone (for fixed lower limits of the 
integrals (\ref{Q-def},\ref{w-def})) labels independent 
classical solutions, as seen 
e.g. from the classification from their global properties 
\cite{kat86,KumTie}. 
If matter is added the (classical) conservation 
law generalizes to $\nabla_{\mu}({\cal{C}}+{\cal{C}}^{(m)})=0$   
\cite{kumwidPR,strklo-CQG} which 
will not be directly relevant for our present work.

In terms of (\ref{B1-0} - \ref{B3-0}) 
${\cal C}_0$ of (\ref{delta-C-b}) is evaluated 
easily. Assuming fixed lower limits of all integrations 
as in (\ref{Q-def}, \ref{w-def}) we may 
rewrite (\ref{T-0}) with a homogeneous solution $\bar{t}$ 
($\nabla_0 \bar t = 0$) as
\begin{equation}
\plabel{T-02}
T^{(0)}= \int^{B_1^{(0)}} U(y)dy = Q^{(0)} +\bar{t}   
\end{equation}
where $Q$ has been defined already in (\ref{Q-def}). Employing the same 
trick as going from (\ref{T-0}) to (\ref{T-02}) for the last term in 
(\ref{B3-0}) produces 
$w(p_1)$ of (\ref{w-def}), up to a function $\bar w$ $(\nabla_0\bar w = 0)$, 
i.e. 
\begin{equation}
\plabel{barw}
\nabla^{-1}_0 \left(e^{T^{(0)}}V^{(0)} \bar{p}_2^{(0)} \right) 
= w^{(0)}(p_1) + \bar{w}
\quad .
\end{equation}
This we could also have used to simplify (\ref{B3-0}).
Therefore the result for 
\begin{equation}
\plabel{Cbar}
{\cal{C}}_0=e^{-\bar{t}} \bar{p}_2 \bar{p}_3 -\bar{w} 
\end{equation}
because of (\ref{C-const}), i.e. because of other e.o.m-s, must actually 
be a constant. 
The factor $e^{-\bar t}$ may be absorbed by a redefinition of
$\bar{p}_3$, because it does not contribute to the second term of
(\ref{B3-0}). Of course, $\bar t$ and $\bar w$ may also be dropped
as long as the lower limits in (\ref{w-def}) and (\ref{T-02})
have not been fixed beforehand.

Instead of solving (\ref{0-variation-1}-\ref{xvariation-3}) directly
it is much more convenient to solve the e.o.m.-s from (\ref{lfirst})
and (\ref{lfirstb}) in the formalism of exterior derivatives. By a
trivial generalization of the steps in \cite{str94} and especially in
\cite{kumwidPR} the general solutions for $q_i$ 
are obtained from
\begin{eqnarray}
\plabel{e+form}
e^+&=&e^{Q(X)}X^+df \\
\plabel{e-form}
e^-&=&\frac{dX}{X^+}+X^-e^{Q(X)}df \\
\plabel{omega-form}
\omega&=&-\frac{dX^+}{X^+} + {\mathcal{V}}\, e^{Q(X)}df \\
X^+X^-e^{Q(X)}&=&{\cal{C}}_0-w(X)
\end{eqnarray}
by introducing the specific 
gauge (\ref{gauge-fix}) for the $0$-components
of the $1$-forms. Identifying the $p_i$ and $q_i$ 
according to (\ref{abbrev}) and introducing
new arbitrary functions $\bar{q}_i$ $((\nabla_0 \bar{q}_i)=0)$
one obtains
\begin{eqnarray}
\plabel{e+form1}
e_1^+&=&\bar{q}_3e^{Q^{(0)}} \\
\plabel{e-form1}
e_1^-&=&\bar{q}_2 
- \frac{\bar{q}_3}{\bar{p}_2^2}\left(w^{(0)}+\bar{w}\right)
+\int_{x'}{\bar{q}'}_1 \nabla^{-1}_{0xx'}           \\
\plabel{omega-form1}
\omega_1&=&\bar{q}_1+\frac{\bar{q}_3}{\bar{p}_2}V^{(0)}e^{Q^{(0)}}-
\frac{\bar{q}_3}{\bar{p}_2}U^{(0)}\left(w^{(0)}+\bar{w}\right)+
\bar{p}_3\bar{q}_3e^{-\bar{t}}U^{(0)}
\end{eqnarray}
where $Q^{(0)}$ etc. are defined as in (\ref{T-0}) and (\ref{Q-def}).
The new arbitrary functions $\bar q_i$ are restricted by
\begin{eqnarray}
\plabel{func-rel1}
\nabla_1\bar{p}_1+\bar{p}_3\bar{q}_3e^{-\bar{t}}-\bar{p}_2\bar{q}_2&=&0 \\
\plabel{func-rel2}
\nabla_1\bar{p}_2+\bar{q}_1\bar{p}_2 &=&0 
\end{eqnarray}
and (\ref{Cbar}). It can be easily verified that (\ref{func-rel2})
follows by inserting the solutions for $p_i$ and $q_i$ 
into (\ref{0-variation-3}), similarly (\ref{func-rel1}) is a
consequence of (\ref{0-variation-2}), if (\ref{func-rel2}) is used
again. (\ref{Cbar}) corresponds to a linear combination of 
(\ref{0-variation-1},\ref{0-variation-2},\ref{0-variation-3}), as indicated
already above.

The gauge (\ref{gauge-fix}) leaves undetermined
residual transformations. One 
derives easily that local Lorentz boosts $\gamma = \bar{\gamma} 
(x)$ ($\nabla_0 \bar{\gamma} = 0$) 
\begin{eqnarray}
\plabel{e1+}
e_1^+ &=& q_3 \to e^{\bar{\gamma}} q_3 \\
\plabel{e1-}
e_1^- &=& q_2 \to e^{-\bar{\gamma}} q_2 \\
\plabel{w1}
\omega_1 &=& q_1 \to q_1 - {\nabla}_1 \bar{\gamma}
\end{eqnarray}
and $\bar{\gamma}$-dependent diffeomorphisms 
($\nabla_0\bar{x}^\mu=0$) 

\begin{eqnarray}
\plabel{x0}
x^0(x') &=& \bar{x}^0(x') + {\nabla}_0^{-1}
e^{\bar{\gamma}(x')} \\
\plabel{x1}
x^1(x') &=& \bar{x}^1(x')
\end{eqnarray}
still may be used to simplify the solutions.

Applying the Lorentz boost
(\ref{e1-}) to $\bar{p}_2$ this quantity can be fixed as $\bar{p}_2=\pm 1$.
Eq.\ (\ref{x0}) enables us to make $\bar{p}_1=0$. Removing the
regularization ($\tilde{\mu}\to 0$) in the manner described in the
Appendix B at the same time requires some care. For small coordinate
values $x^0$ the combination $\bar{p}_1+\nabla_0^{-1}\bar{p}_2$ in this
residual gauge simply becomes $x^0$. 

The  component $g_{11}$ of the metric 
coincides with the Killing norm in the EF gauge. It is expressed 
as $g_{11}=2e_1^+e_1^-=2q_2q_3$ in
terms of (\ref{e+form1}), (\ref{e-form1}).
In this special gauge from (\ref{func-rel2}) $\bar{q}_1=0$ and with
(\ref{func-rel1}) the usual EF-form of the line element in terms of
coordinates $x^0$ and  $x^1$ 
\begin{equation}
ds^2=2q_3dx^1\left( dx^0 + \left[ {\cal{C}}_0-w(x^0)\right]dx^1\right)
\end{equation}
is obtained.

\subsection{``Quantum'' Equations of Motion}

The triple of equations (\ref{1-variation-1} - 
\ref{1-variation-3})
can also be verified to be an immediate
consequence of the generating functional
\begin{equation}
\plabel{W-0jJ}
W^{(0)}(j,J)=\int(d^3q)(d^3p)\det F
e^{i\int_x \dot{q}_ip_i+q_1p_2-q_3\mathcal{V}(p_1,p_2p_3)+j_iq_i+J_ip_i}
\end{equation}
which follows in the matterless case from (\ref{L-eff}) 
and (\ref{L-source}) where terms with $S$ and $P$ and the 
improved related measure are dropped.
Integrating $(d^3q)$ and $(d^3p)$ leads to 
\begin{eqnarray}
\plabel{55kum}
W^{(0)}&=&\exp \frac{i}{\hbar}\int
\left[\hbar J_iB_i+{\cal{L}}^{HK} \right] d^2x \\
\plabel{56kum}
{\cal{L}}^{HK}&=&\tilde{g}e^T \left(\hbar j_3 -V\right)
\end{eqnarray}

The ``quantum'' e.o.m.s for the $p_i$ are obtained by varying the $q_i$
in $(d^3q_i)$. In this way e.g. for $\delta q_1$ a relation
\begin{equation}
\delta W^{(0)}=0=i\int(d^3q)(d^3p)[(\ref{0-variation-1})+
\mbox{terms with $j,J$}]
\det F e^{i\int_{x'}...}
\end{equation}
follows, where (\ref{0-variation-1}) means the left hand side of 
(\ref{0-variation-1}). Replacing 
$p_i \to \frac1i \frac{\delta}{\delta J_i}$ the square bracket may be
pulled outside the integral which can be evaluated as before.
At $J_i=j_i=0$ we thus obtain (apart from a factor $W^{(0)}$) 
eq.\  
(\ref{1-variation-1}) with $p_i$ replaced by $B_i^{(0)}$. In a similar
manner from $\delta q_2$ and $\delta q_3$ eqs. (\ref{1-variation-2})
and (\ref{1-variation-3}) follow with the same replacements. Because 
$B_i^{(0)}$ indeed are the solutions of the classical equations their
``quantum'' versions are fulfilled identically, in other words the
``expectation values''
\begin{equation}
\langle p_i \rangle= B_i^{(0)}
\end{equation}
coincide with classical solutions (for vanishing external sources). 
In order to obtain the quantum version of (\ref{delta-C-b}) we may
proceed as in equations (\ref{1-variation-1}-\ref{1-variation-3})
of which (\ref{delta-C}) has been 
a consequence $(p_i \to \frac1i \frac{\delta}{\delta J_i} \to B_i^{(0)})$:
\begin{equation}
{\cal{C}}=e^{Q(B_1(0))}B_2^{(0)}B_3^{(0)} +w(B_1^{(0)})
\end{equation}
with $\nabla_0 {\cal{C}}=0$. 
In the classical case $\nabla_1 {\cal{C}}=0$ is a consequence of relations
(\ref{0-variation-1}-\ref{0-variation-3}).
These relations are 
nothing else but the constraints appearing in the Hamiltonian 
\cite{kum97}. They clearly do not follow directly from 
the gauge fixed action but are related to Slavnov-Taylor like
identities.
These are obtained by the available gauge
transformations (local Lorentz transformations 
$\delta e_1^{\pm}=\pm \delta \gamma (x) e_1^{\pm}$ etc. and
diffeomorphisms $\delta \zeta^{\mu}(x))$ in the variables of the path 
integral (\ref{W-0jJ}). A straightforward computation shows that
$\delta \gamma (x)$ indeed produces the Lorentz constraint 
(\ref{0-variation-1}), $\delta \zeta^0$ yields (\ref{0-variation-2}),
whereas for $\delta \zeta^1$ the identity $q_i\partial_1 p_i=0$
follows which is easily verified from 
(\ref{0-variation-1}), (\ref{0-variation-2}) and (\ref{0-variation-3}).

Thus the expectation values of $q_i$ 
\begin{equation}
\langle q_i \rangle = \left. \frac{1}{W^{(0)}}
\frac{\delta W^{(0)}}{i \delta j_i}\right|_{J=j=0}=
\left. \frac{\delta }{i \delta j_i}\int {\cal{L}}^{HK}d^2x
\right|_{j=0}
\end{equation}
must be identical to the
classical solutions (\ref{e+form1})-(\ref{omega-form1}).
We re-emphasize the importance of ${\cal{L}}^{HK}$ especially for the
present case.

Of course, $\nabla_1{\cal{C}}=0$ then also is true ``quantum
mechanically''.

According to the quantum point of view the
values of ${\cal{C}}_0$ must be related to a superselection rule. The
``full'' generating functional (${\cal{C}}$ now runs through all
constant values) reads
\begin{equation}
\plabel{kum73}
W^{\mbox{\scriptsize total}}=\int d {\cal{C}} W^{(0)} ({\cal{C}},J,j) .
\end{equation}
But each ``expectation value'' of an operator 
${\cal{O}}_{{\cal{C}}_0}\delta ({\cal{C}}_0-{\cal{C}})$ related
to  each superselection sector with fixed ${\cal{C}}_0$ 
leads to the use of just
the $W^{(0)} ({\cal{C}}_0,J,j)$ as discussed above.

In view of the general nature of this argument, it should be
possible to apply it to the case with matter ($S$-fields)
as well. But at the classical level already the determination of the
matter contribution ${\cal{C}}^{(m)}$ to ${\cal{C}}$ is nontrivial,
because that contribution (in contrast to
${\cal{C}}={\cal{C}}^{\mbox{\scriptsize (geom)}}$
treated in this section) is nonlocal and in general cannot be obtained
explicitely by solving the equations of motion \cite{KumTie}. 
On the other hand, the perturbative theory in terms 
of scalar loops described below can be conjectured to take care 
of this order by order if matter fields are retained in eq.\ 
(\ref{W-0jJ}).

As a consequence of 
the preceding comparison between the classical and the
``quantum'' version for the matterless case we also see that the first
order form of the action (\ref{lfirst}) -- which has been crucial
to obtain a simple solution for the geometric part of all 2d covariant
theories -- in the quantum case quite naturally suggests the
introduction of sources $J_i$ for the momenta. Of course, for the
subsequent computation of correlation functions of ``physical'' fields
we always have $J_i=0$. However, those sources remain a crucial 
tool for a simple formulation of the central conservation law
(\ref{delta-C-b}). 

We conclude this section with an apology to the reader that we found
it necessary to deal in considerable detail with the equivalence
between a ``quantum'' and a classical formulation at all. However, we
feel that it is a somewhat unusual situation indeed to obtain a
basically classical solution from an (exact) generating functional.
Using this section as a basis we will
also have to continue below at the quantum level when we proceed to
take interactions with scalar matter into account, where the 
scalar fields induce loop
corrections to the exact geometric part.

\section{Integration of Scalars}
\plabel{integration}

\subsection{Generating functional}

For a perturbation theory in terms of the 
scalar field with exact geometric 
integrations it is sufficient to collect 
systematically the terms of ${\cal O}(S^2)$  and perform a 
Gaussian integration in order to obtain the propagator of $S$. 
Terms $O(S^{2(n+1)}),\; n > 0$ yield the interaction Lagrangian 
${\cal L}_{\mbox{\scriptsize int}}^{(S)}$ (cf.\ (\ref{L4-eff})  
below). 

We first treat approach a).
For $\hat B_1$ and $\hat B_2$ the terms ${\cal O}(S^2)$ can be 
read off from (\ref{B12},\ref{B22}). 
For $\hat B_3$ they can be summarized in the 
nonlocal expression $H_{xy}$ as 
\begin{equation}
\plabel{B3-hat}
\hat{B}_3 = B_3 +\int_{y} H_{xy} \left({\partial_0}S(y) \right)^2
\end{equation}
where $B_3$ is defined in (\ref{B32}).
 From the definition of $\hat T = T(\hat B_1,\hat B_2)$ one computes 
\begin{eqnarray}
\plabel{T-hat2}
\hat{T}_x &=& T(B_1,B_2)_x - \int_{y} G_{xy}\left({\partial_0}S(y)\right)^2 \\
\plabel{Gxx}
G_{xy} &=&
\int_{z}\nabla_{0xz}^{-1}({U_z}'B_{2z}\nabla_{0zy}^{-2}
+U_z\nabla_{0zy}^{-1})
\end{eqnarray}
The prime in $U$ denotes a differentiation with respect to
the argument. 
With
\begin{equation}
\plabel{Vhat}
\hat{V}_x=V(\hat{B}_1)_x=V(B_1)_x-{V_x}'
\int_{y} \nabla_{0xy}^{-2}\left({\partial_0}S(y)\right)^2
\end{equation}
one arrives at

\begin{equation}
\plabel{Hxx}
H_{xy}=e^{-T_x} \int_{z}\nabla_{0yz}^{-1}e^{T_z}
\left[\left(G_{zy}-G_{xy}\right)\left(\hbar j_3-V\right)_z
+{V_y}'\nabla_{0zy}^{-2} \right] -\bar{p}_3e^{-T_x}G_{xy} \quad .
\end{equation}
Together with
an analogous expansion of $\hat{\cal L}^{HK}$ for
${\cal L}_{(3)}^{\mbox{\scriptsize eff}}$ in (\ref{Wtilde}) this yields
\begin{eqnarray}
\plabel{L4-eff}
\frac{{\cal L}_{(4)}^{\mbox{\scriptsize{eff}}}}{\hbar}&=&
J_iB_i
+\frac{\tilde{g}e^T}{\hbar}\left(\hbar j_3 -V\right) + \\
{}&+&\frac{1}{\hbar} 
\left((\partial_0 S)(\partial_1 S)-E_1^-(\partial_0 S)^2 \right) +SQ 
+ \frac{{\cal L}^{(S)}_{\mbox{\scriptsize int}}}{\hbar}
\quad .
\nonumber
\end{eqnarray}
The abbreviation
\begin{eqnarray}
\plabel{E1-}
E_1^-(x) &=& -\int_{y} [ \hbar 
(J_1(y)\nabla_{0yx}^{-2} +J_2(y)\nabla_{0yx}^{-1} +
  J_3(y)H_{yx} )+
\\
\ && +\tilde {g}[e^{T(x')}  (\hbar j_3 - V)]_yG_{yx} 
-{V'}_y\nabla_{0yx}^{-2}) ] \quad .
\end{eqnarray}
indicates the role of this quantity, replacing $e_1^-$ in the EF-gauge
version of the term $\sqrt{-g}S\square S$.
The $S$-integration to be performed now becomes
\begin{equation}
\plabel{kum96}
\widetilde W =
\int (dS)\sqrt{\det f}\; \exp\, \int_x\,\frac{i}{\hbar}\, \left( 
-\frac{1}{2}\, \left[ S \sqrt{-g}\, \square S \right]_{EF} + SQ +  
{\cal L}^{(S)}_{\mbox{\scriptsize int}} 
\right) 
\end{equation}
where it should be noted that according to (\ref{L4-eff}) in the
EF gauge the quadratic expression in $S$ is independent of $e_1^+$,
with only $e_1^- = E_1^-$ determining the ``background'' for that
integral. $e_1^+ = E_1^+ = f$ only enters through the measure $\sqrt{\det f}$. Thus
(\ref{kum96}) corresponds to the standard Polyakov integral with the
metric determined by  $E_1^-$ from (\ref{E1-}) and $E_1^+=f$:
\begin{equation}
\plabel{kum97}
g_{\mu \nu}=f\begin{pmatrix}0&1 \\ 1 & 2E_1^- \end{pmatrix}
\end{equation}
Completing the square in $S$ the Gaussian integral  
(\ref{kum96}) yields the 
Polyakov term ${\cal L}_P$ \cite{polyakov}  and the 
propagator $\vartriangle_{xy} = (f(x)^{-1/2)} \square^{-1}_{xy} 
f(y)^{-1/2}$  for  $S$:

\begin{eqnarray}
&&\widetilde W=\exp \left[ i\int_x 
\left[\frac{\hbar}{2}\left(\int_{y}Q_x\vartriangle_{xy}^{-1}Q_y
\right)^{EF} 
+\frac{{\cal L}_P^{EF}}{\hbar}+J_iB_i +\right. \right.
\nonumber \\
&&\qquad\qquad\qquad\qquad
\left. \left. +\frac{\tilde{g}e^T}{\hbar}
\left(\hbar j_3 -V(B_1)\right)+ \frac{{\cal{L}}^{(S)}_{\mbox{\scriptsize 
int}}}{\hbar}\right]\right]  
\plabel{mainW}
\end{eqnarray}
In (\ref{mainW}) EF means that the Polyakov action
\begin{equation}
\plabel{L-pol}
{\cal L}_P =-\frac{\hbar}{96\pi}
\int_x\int_{y}\sqrt{-g}R_x\square_{xy}^{-1}R_y  \quad .
\end{equation}
is to be understood to depend on the (source-dependent!) metric (\ref{kum97}):
\begin{eqnarray}
&&{\cal L}_P(E_1^+,E_1^-)=\frac{-\hbar}{96\pi}\int_x \int_y
(\partial_0^2 E_1^- -\Gamma \ln E_1^+)_x\Gamma^{-1}_{xy} 
(\partial_0^2 E_1^- -\Gamma \ln E_1^+)_y\nonumber \\
&&\qquad \Gamma = \partial_1\partial_0 -\partial_0 E_1^-
\partial_0 \plabel{Lpol-gauge}
\end{eqnarray}
It obviously does not possess the simple form of the conformal gauge.
The factorization of the contribution from the Polyakov term in
(\ref{mainW}) means that this one-loop contribution from the scalars
appears disconnected from the propagator part of ${\cal O}(Q^2)$. 
(\ref{mainW}) is a function of external 
sources and $f$. But $f$ cancels in the propagator and thus
resides in the Polyakov term only. We may
rewrite (\ref{mainW}) as
\begin{equation}
\plabel{d101}
\widetilde W(f,j,J,Q)=
\exp \left( \frac i\hbar {\cal L}^{(S)}_{int}\left( \frac 1i 
\frac \delta{\delta Q} \right) \right) \widetilde{W}_{prop}
\widetilde{W}_0(j,J) \exp \left[ \frac i\hbar \int {\cal L}_P^{EF}
\right]
\ .
\end{equation}
We recall that 
${\cal L}_{int}^{(S)}$ is obtained from (\ref{Wtilde-b}) by
keeping the terms of order $S^4$ and higher, and
$S$ is replaced by the functional 
derivative $\frac 1i \frac \delta{\delta Q}$.
The other factors in (\ref{d101}) are
\begin{eqnarray}
\plabel{kum102}
-i\ln \widetilde{W}_0 (j,J)&=&\int \left[ J_iB_i +
\frac{\tilde{g}}{\hbar}e^{T(B_1)}
(\hbar j_3 -V(B_1)) \right] \\
\plabel{kum104}
-i\ln \widetilde{W}_{\mbox{\scriptsize prop}}(j,J,Q)&=&
\frac{\hbar}{2}\int_{x}\int_{y}Q_{x}\vartriangle_{xy}^{-1}Q_{y}
\end{eqnarray}
As a final step in approach a) $\widetilde W$ has to be integrated with a
$\delta$-function as in (\ref{tilS}),
\begin{equation}
\plabel{again} 
W=\int(df)\delta(f-\frac{\delta}{i\delta j_3})\widetilde{W}
=\int(df)\delta(f-\frac{1}{i \widetilde{W}}
\frac{\delta \widetilde{W}}{\delta j_3})\ .
\end{equation}
Using (\ref{d101}) with (\ref{kum102}) we may write
\begin{equation}
\frac{1}{i \widetilde{W}}
\frac{\delta \widetilde{W}}{\delta j_3}=f^{(0)}+\hbar
Y(f,\mbox{sources})
\end{equation}
where 
\begin{equation}
\plabel{f-0}
f^{(0)}=\tilde{g}e^{T(B_1,B_2)}
\end{equation}
represents the ``background'' value of $q_3=e_1^+$ consisting of the
classical background together with sources $j_1$, $j_2$ for $e_1^-$
and $\omega_1$ in $B_1$, $B_2$. $\hbar Y$ is the remainder, being of
higher order in $\hbar$. The $\delta$-function in (\ref{again}) is
solved by some $f=\hat f$ obeying 
\begin{equation}
\plabel{fY}
f=f^{(0)} + \hbar Y( f,\mbox{sources})\ .
\end{equation}
In general no exact solution of this equation is available. Thus an
iterative solution must be sought. The most obvious one would consist
in an expansion in $\hbar$. To lowest order in $\hbar$ we would 
have 
$ f=f^{(0)}+\hbar f^{(1)}+...$ with $f^{(1)}=Y(f^{(0)},...)$. Possibly
one may be able to do better by including some part of $Y$ already to
lowest order. This would correspond in spirit to the semiclassical
approach which uses the Polyakov action as a contribution to the
classical one (cf.\ e.g.\ the fourth and fifth reference
in \cite{man91}).
When such a (approximate) solution $f=\hat f$ to the vanishing 
argument of the $\delta$-function in 
(\ref{again}) has been
obtained it yields   
\begin{eqnarray}
\ln W &=& -\ln \det \left[\delta_{xx'}-
\left(\frac{\delta}{\delta f(x')} \frac{1}{i\tilde W} 
\frac{\delta \tilde W}{\delta j_3(x)}\right)_{f=\hat f} \right] \\
{}&=& -\int_x \ln 
\left[1-\left(\frac{\delta}{\delta f(x)} \frac{1}{i\tilde W} 
\frac{\delta \tilde W}{\delta j_3(x)}\right)_{f=\hat f}\right]
\label{eq7}
\end{eqnarray}
This is true whether $\hat f$ is known exactly or  perturbatively. Of
course, in a perturbative expansion in $\hbar$ (or in some other
``small'' parameter) the determinant (or the log) can be expanded as
well. E.g. to lowest nontrivial (first) order in $\hbar$ (\ref{eq7})
becomes 
\begin{equation}
\ln W \equiv \left(-\int_x \frac{1}{\tilde W^2}
\frac{\delta \tilde W}{\delta f(x)}\frac1i 
\frac{\delta \tilde W}{\delta j_3(x)}+
\int_x\frac{1}{\tilde W}\frac{\delta^2 \tilde W}{\delta j_3(x)\delta f(x)}
\right)_{f=f^{(0)}}
\label{eq8}
\end{equation}
The first term in (\ref{eq8}) can be interpreted as the coupling of
the external field $f^{(0)}$ to $e_1^+$. The second one is a
``self-loop'' contribution.

Clearly the necessity of an iterative (perturbative) solution
for $f$ in approach a) conceals our program to start from an exact
solution of the geometric part:  It is apriori unclear whether 
this expansion is related to the one in scalar loops. 
Therefore, we now turn to approach b). 
The generating functional (\ref{B32-b})
after integrating over $q_i$ and $p_i$ becomes
\begin{equation}
\widetilde{\widetilde W} =\int (dS) \exp \frac i\hbar
\int d^2x \widetilde{\widetilde {\cal L}}^{eff}_{(3)}
\plabel{b105}
\end{equation}
where
\begin{equation}
\widetilde{\widetilde {\cal L}}^{eff}_{(3)}=
\hbar (J_1\hat B_1 +J_2\hat B_2 +J_3\widetilde{\widetilde B}_3
+SQ)+(\partial_0 S)(\partial_1 S) +\widetilde{\widetilde {\cal L}}^{HK}
\plabel{b106/1}
\end{equation}
According to (\ref{B32-b}) and (\ref{L-HK-b}) the ghost contribution
$\tilde{\tilde l}$ from $\widetilde{\widetilde B}_3$ as well as from
$\widetilde{\widetilde {\cal L}}^{HK}$ still appear {\it linearly}.
Therefore, the identity (\ref{b19}) may be simply used backwards:
\begin{equation}
\widetilde{\widetilde W}=\int (dS) \sqrt{\det E_1^+}
\exp \frac i\hbar
\int d^x  {\cal L}^{eff}_{(3)}
\plabel{b106/2}
\end{equation}
${\cal L}^{eff}_{(3)}$ is precisely the expression (\ref{Wtilde-b})
again and in the measure 
\begin{equation}
E_1^+=\hbar e^{-\hat T}\nabla_0^{-1} e^{\hat T} J_3+
\tilde g e^{\hat T}
\plabel{b106/3}
\end{equation}
even at $J_3=0$ shows a dependence on the scalar field $S$. Thus 
$S$ itself influences the measure in the path integral and  
backreaction is fully taken into account. Of course,
(\ref{b106/2}) with (\ref{b106/3}) cannot be integrated exactly.
But we are interested in a loop expansion for $S$ only.
Therefore, from the determinant in (\ref{b106/2}) as well as from
${\cal L}^{eff}_{(3)}$ terms $O(S^2)$ will contribute together
 to  a Polyakov type action and to a propagator for $S$. Again 
higher order terms $O(S^{2n})$, $n>2$, may be collected in an
interaction Lagrangian where $S$ is replaced by $\frac 1i
\delta /\delta Q$ as in (\ref{d101}).

The expansion for ${\cal L}^{eff}_{(3)}$ has been given already 
above. The additional contributions from $\sqrt{\det E_1^+}$
are determined by (for simplicity we restrict $J_3 = 0$) in   
(\ref{b106/3})  
\begin{equation}
\sqrt{\det E_1^+} = \sqrt{\det e^T} \exp 
\left[ -\frac i2 \int_x \int_y 
G_{yx} (\partial_0 S)^2_x +\dots \right]
\plabel{b106/4}
\end{equation}
Thus the classical background enters -- beside the sources $j_i$ 
(cf.\ eqs.\ (\ref{T-hat2}) and (\ref{Gxx})) -- as expected, to lowest
loop order in the proper definition of the $S$-measure.
On the other hand, for the generalized Polyakov term
(with $E_1^+=e^{T}$), for the propagator of the scalar
$S$ and for scalar vertices additional terms will emerge.

The difference between the $S^2$ terms 
in the two approaches a) and b) comes
from the exponential (\ref{b106/4}). The contribution from
(\ref{b106/4}) is of order $\hbar^0$, compared to the order
$\hbar^{-1}$ contribution from ${\cal L}^{eff}_3$. In approach a)
a term like (\ref{b106/4}) appears as a one-loop effect
(tadpole).
To lowest order in $\hbar$, which is considered below,
the propagators for the scalar coincide in both approaches. 
Comparing the two approaches, 
b) clearly has the advantage of being formally exact,
avoiding the iterative solution in the geometric variable
$e_1^+=\sqrt{-g}$ of (\ref{fY}) whose relation to the expansion 
in scalar loops is not obvious. 
We must admit though that there is always a danger involved in manipulating
the measure for the $S$-integral. One verifies easily that the formula
(\ref{Lpol-gauge})
for the Polyakov action from the $S$-integral is not consistent with
arbitrarily extracting (part of) the (background) measure before the
integral is done. Therefore case b)
and the situation with a general
background field $f$ (case a))  
superficially seem to be different. Another important remark
concerns a necessary ultra violet regularization of the scalar
propagator.  In approach a) it sometimes may be convenient to use an 
$f$-dependent regularization. For
example, this might be a large eigenvalue cut off for an $f$-dependent
differential operator. Hence, though the first two terms in
(\ref{d101}) formally do not depend on $f$ one cannot pull them out of
the $f$-integral.

\section{Effective scalar theory}
\label{sub-eff}
After integrating out all geometric degrees of freedom we are
left with an effective theory of the scalar field $S$ with
non-local self-interaction.
In this section we take a step back from our 
somewhat involved quantum expressions and show
that again our lowest order contributions reproduce their classical
counterparts. We also derive the effective $S^4$ vertices for 
spherically reduced gravity (SRG).

\subsection{Effective propagator of scalars}

The propagator 
$\left[ \sqrt{-g}\square \right]_{xx'}^{-1}=\vartriangle_{xx'}^{-1}$ in
(\ref{kum104}) only depends on $E_1^-$. 
An exact evaluation (to this order in the matter 
fields!) still requires the solution of
\begin{equation}
\plabel{inv-quabla}
2\partial_0 \left(\partial_1 -E_1^-\partial_0\right)
\vartriangle^{-1}_{xx'}=\delta^2(x-x')
\end{equation}
which may be reduced to finding the inverse of $\vartheta = 
\partial_1 - E_1^- \partial_0$ in

\begin{equation}
\plabel{vartheta}
\vartriangle^{-1}_{xx'}=\frac{1}{2}\int_{x''}\vartheta^{-1}_{xx''} 
\nabla_{0x''x'}^{-1} \quad .
\end{equation}
As noted in \cite{nonpert} we may write

\begin{eqnarray}
\plabel{inv-theta}
\vartheta_{xx''}^{-1} &=& P^{-1}\nabla_1^{-1}P \\
\plabel{p-order}
P &=& {\cal P} \exp \left(-\int_{x'}
\nabla_{1xx'}E_1^-({x'}^1,x^0)\partial_0 \right)
\end{eqnarray}
where P contains the path ordering $\cal P$ and is local in the 
overall variable $x^1$. 

Whenever $W$ is used to calculate 
correlation functions after (functional) differentiations with 
respect to the sources in $E_1^-$, those sources are set to zero. 
Therefore, the relevant $E_1^-$ for such computations is
($\tilde{g}'=\bar{g}+ {\cal O}(\mu)$ in the second term)
\begin{equation}
\plabel{E1-rel}
E_1^- \mid_{j=J=0} = E_1^{-(0)}=- \int_{y}
\tilde{g} e^{{T^{(0)}}_y}({V^{(0)}}' +U^{(0)}V^{(0)})_y 
\nabla_{0yx}^{-2}
\ .
\end{equation}
In (\ref{E1-rel}) the index ($0$) indicates a dependence on the 
``classical'' solution $B_1^{(0)}, B_2^{(0)}$ of section 3. The 
first term on the r.h.s.\ of (\ref{E1-rel}) then indeed 
(after the integrations implied by $\nabla_{0xx'}^{-2}$) coincides with the 
classical solution of $q_2 = e_1^-$.

Higher order scalar vertices can be obtained in the same way by
straightforward calculations.

\subsection{Effective scalar interaction in spherically
reduced gravity}
Certainly the most interesting case among 2D dilaton gravities
is SRG, which corresponds to
$U_{SRG}(X)=-(2X)^{-1}$ and $V_{SRG}(X)=-2$. Here the previous
formulae for the effective interaction vertices must be modified.
Due to the singularity in $U(X)$ at $X=0$
the right hand side of (\ref{E1-rel}) become divergent. 
One may try to introduce
definitions of inverse derivatives valid on the half line
${\rm R}_+$ (see, e.g.\ \cite{LiMi}). This is not likely to work
either, because higher vertices involve higher derivatives
of $U$, which are more and more singular at $X=0$ and cannot
be made square integrable with any reasonable weight function.
However, as we demonstrate below, all integration constants in 
$\nabla_0^{-1}$ can be recovered if one takes scalar fields
to be localized at certain points $x^0_1,\dots\ ,x^0_n$ in
the ``time'' variable and compares the geometric fields
in the ``past'' with known empty space classical solutions.
Thus a kind of causality condition will be used.

We restrict ourselves to approach a). Since this approach
does not mix different orders in $\hbar$, it is easier to
handle in the present context. We consider first the general
case before specializing to SRG.

The effective vertex of order $2(n+1)$ has the form:
\begin{equation}
\int dx_1 \dots dx_{n+1} {\cal S}^{2(n+1)}(x_1,\dots ,x_{n+1})
(\partial_0 S)^2(x_1) \dots (\partial_0 S)^2(x_{n+1})
\label{a88}
\end{equation}
We consider the vertex for vanishing sources $j_i$ only. Then in 
approach a) ${\cal S}^{2(n+1)}$ is given by the
$(n+1)$th functional derivative of ${\cal L}^{HK}$ (\ref{L-HK}) 
with respect to $j_2$ because $(\partial_0 S)^2$ enters in the
combination $[\hbar j_2-(\partial_0 S)^2]$:
\begin{equation}
{\cal S}^{2(n+1)}=(-\hbar)^{-n-1}\frac 1{(n+1)!}
\frac {\delta^{n+1}}{\delta j_2^{n+1}} {\cal L}^{HK}\vert_{j=0}
\label{b88}
\end{equation}
For vanishing sources $j_1=j_3=0$,  $T$ in ${\cal L}^{HK}$ can 
be expressed in a similar way as (\ref{T-02})
\begin{equation}
T=\int_{y_0}^{B_1}U(y) dy \ , \label{T88}
\end{equation}
where $B_1 = B_1(j_2)$ only and thus according to  
(\ref{B12}), (\ref{B22}) 
$\nabla_0 B_1=B_2$. Here the arbitrariness in the 
choice of the inverse derivative $\nabla_0^{-1}$
in (\ref{That})  is taken into account by an, at first,
arbitrary lower limit
of integration. In the limit of a vanishing $\mu$, the IR 
regulator, $y_0$ can depend on $x^1$ only, and, therefore,
does not depend on sources. 
For simplicity, we put $\tilde g=1$. Let us evaluate the first derivative
in (\ref{b88}) explicitly:
\begin{eqnarray}
&&{\cal S}^{2(n+1)}=(-\hbar )^{-n}\frac 1{(n+1)!}
\frac {\delta^{n}}{\delta j_2^{n}} E^-_1\vert_{j=0} \nonumber \\
&&E_1^-=-\nabla_0^{-2} (e^T(V'+UV))
\label{c88}
\end{eqnarray}

As a next step, we embed the  $E^-_1$ defined in (\ref{c88})
into a  classical system described by the equations of motion 
(\ref{1-variation-1})-(\ref{1-variation-3}), (\ref{xvariation-1})-
(\ref{xvariation-3})  in the presence of an
external source $j_2$. We do not need the constraint
equations (\ref{0-variation-1})-(\ref{0-variation-3}) here. In the
presence of $j_2$ only the equation (\ref{1-variation-2}) is
modified:
\begin{equation}
\dot p_2=\hbar j_2 \label{d88}
\end{equation}
We simply assume that the equations 
(\ref{1-variation-1})-(\ref{1-variation-3}) are satisfied
and use them to define the functions $p_1$, $p_2$ and $p_3$. Note,
that these equations were found to hold if they are interpreted as
equations for expectation values of corresponding quantum fields
in the presence of external sources, i.e.\ only $j_2$ here. 
The rest of the equations of motion (\ref{xvariation-1})-
(\ref{xvariation-3}) yields 
\begin{eqnarray}
&&\partial_0^2 q_2 =-q_3(V'+UV) \label{e88}
\\
&&q_3=e^T ,\nonumber
\end{eqnarray}
where integration constants in the definition of $q_3$ are encoded
again in the lower limit $y_0$ of the integral (\ref{T88}). 
Hence, $E_1^-(j_2)$ can be identified with a classical solution
for $q_2$. Fixing the ambiguity in the definition of $\nabla_0^{-2}$ is
equivalent to fixing integration constants in the classical
field equation (\ref{e88}). As we shall demonstrate below, this
is equivalent to choosing asymptotics of the classical gravitational
background.

To obtain the $n$th functional derivative of $E_1^-$ it is 
sufficient  to
take $j_2$ localized at $n$ different points:
\begin{equation}
j_2(x)=-\sum_{k=1}^n c_k \delta (\bar y_k-x)
\label{f88}
\end{equation}
then we can expand $E_1^-(j_2,x)$ in a power series of $c_k$.
In the resulting sum the
coefficient of the term with 
$(-1)^n\prod_{k=1}^n c_k$ will give the desired
functional derivative. Indeed, consider the expansion of $E_1^-$
\begin{equation}
E_1^-=\sum_n \frac 1{n!} \int dx_1\dots dx_n {\cal E}^n
(x_1,\dots ,x_n)j_2(x_1)\dots j_2(x_n) \ .
\label{e188}
\end{equation}
For $j_2$ given by the equation (\ref{f88}), we have
\begin{equation}
E_1^-(j_2)=(-1)^n\prod_{k=1}^n c_k {\cal E}^n
(\bar y_1,\dots ,\bar y_n)\ +{\rm other\ \ terms}.
\label{ec88}
\end{equation}

We must now fix a particular solution $E_1^-(j_2,x)$. This is
done by some "causality" condition. For $x^0$ in the asymptotic
region, $x^0 >\bar y_k^0,\ k=1,\dots ,n$, the function
$E_1^-(j_2,x)$ must coincide with a fixed vacuum solution
$E_1^-(0,x)$.
Given a vacuum solution $E_1^-(0,x)$, the function $E_1^-(j_2,x)$
is uniquely defined and non-singular.

As demonstrated above and at the end of Appendix B, by using the residual 
gauge freedom and after removing the regularization one can
choose the vacuum values $p_2(j=0,x)=\bar p_2=1$,
$p_1(j=0,x)=x^0$. For $q_3$ we have from (\ref{e88}) for $U_{SRG}$
and $V_{SRG}$
\begin{equation}
q_3=e_1^+=\exp \int_{y_0}^{p_1} \frac {dy}{-2y} =\sqrt{\frac {y_0}{p_1}}
\label{q388}
\end{equation}
where $y_0$ contributes to an irrelevant scale factor. We put $y_0=1$.
In the absence of sources this amounts just to
$e^+_1=\sqrt{1/x^0}$. The equation for $q_2$ yields
\begin{equation}
q_2=e^-_1(0,x)=4\sqrt{x^0} -m_{\infty}+bx^0
\label{h88}
\end{equation}
Nonzero values of $b$ correspond to Rindler coordinates
referring to an uniformly accelerated 
frame. Therefore we set $b=0$. The effective metric now reads
\begin{equation}
ds^2=\frac{2dx^0dx^1}{\sqrt{x^0}} +2\left( 4-\frac{m_\infty}{\sqrt{x^0}}
\right)(dx^1)^2 \ .\label{ds88}
\end{equation}
After a change of the coordinate $z = \sqrt{x^0}$ the metric (\ref{ds88})
becomes
\begin{equation}
ds^2 = 4 dz dx^1+2\left( 4-\frac{m_\infty}{z}
\right)(dx^1)^2 \ ,\label{sch88}
\end{equation}
which, up to trivial numerical factors, coincides with the Schwarzschild
black hole in Eddington--Finkelstein coordinates with the 
'radial' 
variable $z$ and $m_\infty$ proportional to the mass of the black hole.

With $j_2$ defined in (\ref{f88}) the classical
equations of motion have unique solutions with
given asymptotics for large $x^0$:
\begin{eqnarray}
&&p_2(x)=1+\hbar\sum_k c_k \theta (\bar y_k^0 -x^0) \nonumber \\
&&p_1(x)=x^0+\hbar\sum_k c_k (\bar y_k^0 -x^0) \theta (\bar y_k^0 -x^0)
\label{g88}
\end{eqnarray}
The solution for $q_3$ in SRG is given again by (\ref{q388}). Proper
asymptotic behavior is achieved by taking $y_0=1$.

Consider the quartic scalar interaction ($n=1$) in (\ref{a88}) for SRG. 
For $x^0>\bar y^0$ the function
$E_1^-(j_2,x)$ must be a solution of (\ref{e88})
coinciding with (\ref{h88}). For
$x^0<\bar y^0$ we have:
\begin{equation}
E_1^-=\frac 4{(1+\hbar c_1)^2} 
\sqrt {(1+\hbar c_1)x^0-\hbar c_1\bar y^0}
-m_1-a_1x^0 \label{k88}
\end{equation}
where the integration constants $m_1$ and $a_1$ are defined
by the requirement that $E_1^-$ and its first derivative
are continuous at $x^0=\bar y^0$. Since we are interested in
terms which are linear in $c_1$ only, we arrive at
\begin{equation}
m_\infty -m_1=6\hbar c_1\sqrt{\bar y^0} ,\qquad
a_1=-\frac {2\hbar c_1}{\sqrt{\bar y^0}} \label{l88}
\end{equation}
In this linear order in $c_1$ we also have:
\begin{equation}
E_1^-=-\hbar c_1 \left( 6\sqrt{x^0} -6\sqrt{\bar y^0}
+2\left[ \frac{\bar y^0}{\sqrt{x^0}} -
\frac {x^0}{\sqrt{\bar y^0}} \right] \right)
\label{m88}
\end{equation}
This yields the vertex function:
\begin{equation}
{\cal S}^4 (x,\bar y)=\frac 12
\left( 6\sqrt{x^0} -6\sqrt{\bar y^0}
+2\left[ \frac{\bar y^0}{\sqrt{x^0}} -
\frac {x^0}{\sqrt{\bar y^0}} \right] \right)
\theta (\bar y^0 -x^0)
\label{n88}
\end{equation}
The interaction (\ref{n88}) must be
local in the $x^1$ coordinate. 
A corresponding $\delta$-function is not written explicitely
in (\ref{n88}).

Note, that in the presence of matter fields the constraint
equations are modified. Therefore, the equations 
(\ref{0-variation-1}) - (\ref{0-variation-3}) 
with zero right hand sides are not satisfied here.

\section{Summary and Outlook}
\label{summary}

The central result of our present paper is that for all 2d covariant theories
a quantum field theory can be formulated which treats 
the geometric part (Cartan
variables or metric) {\it exactly.} Interactions with matter are included
in a systematic manner by a loop expansion which automatically 
takes into account
backreaction, order by order. The crucial ingredient of our successful approach
is the use of a light cone gauge for Cartan variables which 
amounts to an Eddington-Finkelstein
gauge for the metric.

In that gauge the remaining geometric variables appear linearly in the action
of the path integral. We found that the proper covariant choice of the measure
for the scalar fields for general dilaton theories can also be transformed into
a similar linear contribution, either by introducing an auxiliary scalar field
or by the introduction of ghost fields. Both types of fields are 
later integrated
out again.

This linearity in the geometric variables produces delta-functions for the 
associated
momenta so that the formal path integral in phase space - to zero loop order
in the matter fields - reduces to the classical solution. Matter fields produce
from their Gaussian integral a (generalized) Polyakov action, 
depending on external
sources. Higher order vertices of the scalars can be computed systematically.

Our present work generalizes ref.\cite{nonpert} 
in an essential manner. There we did
not include yet dilaton theories with kinetic term for the dilaton field. But
only the latter theories contain spherically reduced gravity, i.e. the interaction
of black holes with S-wave matter, as a special case.

The main purpose of our paper
has been to set up the general structure of our approach which
perhaps is the first one to develop at least for some sector 
of gravity theories
the general 
consequences of an `` orthodox'' quantum field theory. 
In this respect our basic philosophy shares certain similarities
with the $S$-matrix approach by 't Hooft \cite{GtH}.
We are fully aware
of the fact that the usual 
difficulties of principle for pure quantum gravity (especially 
nonrenormalizability) cannot
be swept under the rug. But here we have a well-defined framework, fixed by
quantum experience in Minkowski space. Any necessary modification 
from the unification
of gravity and quantum theory must show up in its precise relation to orthodox
quantum theory. So several of the many open question 
left by our work may already
have deeper implications. We list just a few:

Performing the path integral for the geometric 
variables we integrated the field
variable $ q_{3}=e_{1}^{+}$ from $- \infty$ to $+ \infty$. But
in the  EF gauge used here this quantity determines the volume element.
Therefore we really integrated over ``negative'' (and vanishing) 
volumes as well. 
Geometrically 
this seems to be very doubtful, but from the point of view of quantum field
theory zweibeine are fields to be varied over their whole range.

In Minkowski quantum field theory path integrals refer to space-time extending
over the whole ${\rm R}^{4}$ . Here we 
find that e.g. for the Schwarzschild black
hole one coordinate is cut off at the singularity given by the 
classical background,
where the latter is not inserted ``by hand'' but appears naturally in this
formalism. This clearly precludes the application of $S$-matrix concepts
to, say, the decay of a given black hole. On the other hand,
 the quantum formation and  disappearance
of some intermediate black hole like object in the scattering matrix element
of two initial scalars into two final ones may be a process to be calculated
in a Minkowski background. In order to obtain theoretical 
information concerning
the eventual self-extinguishing of a black hole by something like 
single quantum
Hawking radiation it may be necessary to develop a sort of ``quantum optics''
for multiquantal states of matter fields. Again our present work should be a
suitable starting point.

It has been noticed some time ago \cite{sch94} 
that a 2d theory of gravity which
allows multiply connected, topologically nontrivial, classical solutions may
carry quantum fluctuations as zero modes in some compact direction. While such
configurations are unlikely in the phase space of SRG they may well play a role
when an additional $ U\left( 1\right)$ gauge field is introduced as well.
{}From technical experience with this case 
\cite{kumwidPR,strklo-CQG} our approach should
basically work as well. But then these additional quantum fluctuations - not
even induced by matter! - will contribute to the geometric part of the path
integral.

The explicit technical machinery has been already 
complicated enough in our present work
for minimally coupled scalar matter. In fact it is quite easy to see that the
basic steps work equally well for SRG with the dilaton field coupled to the
scalars. Only in that case the scalar (nonminimal)
couplings of full SRG are properly taken into account. This
will be, among other things, the object of future work.

These open questions certainly do not form an exhaustive list. In any case we
believe that the range of topics to be explored, starting
 from our present results is not negligible.

\section*{Acknowledgments}

This work has been supported by the Fonds zur F\"orderung der
wissenschaftlichen Forschung project P-12.815-TPH.
One of the authors (D.V.) also thanks the Alexander von Humboldt
Foundation and the 
Russian Foundation
for Fundamental Research, grant 97-01-01186, for financial support.
 
\section*{Appendix A: Background field formalism} \label{AppA}
In this Appendix we rederive
the main result of our previous paper \cite{kum97}
in the framework of a background
field formalism, namely 
that dilaton gravity without matter does not have any loop
effects. We show that a suitable choice of IR regularization all 
anomalous terms are cancelled. Hence the path integral
measure which we are using for the geometric variables is
indeed gauge invariant.

Consider the first order action (\ref{lfirst})
in component notation: 
\begin{eqnarray}
{\cal L}&=&e_1^-(-\partial_0-\omega_0)X^+
+e_0^-( \partial_1+\omega_1)X^+ \nonumber \\
\ &\ &+e_1^+(-\partial_0+\omega_0)X^-
+e_0^+( \partial_1-\omega_1)X^-\nonumber \\
\ &\ &-\omega_1\partial_0X+\omega_0\partial_1X \plabel{aacctt}\\
\ &\ &+(e_0^+e_1^--e_1^+e_0^-)(V(X)+X^+X^-U(X))\; . \nonumber
\end{eqnarray}

Before introducing background fields we determine the gauge symmetries
of the action (\ref{aacctt}).
The simplest way to derive them is to use the canonical formalism.
The canonical coordinates are taken to be
$q_i$
and the corresponding canonical momenta
$p^i$ (cf. (\ref{abbrev})).
The Poisson brackets have the form
\begin{equation}
\plabel{poisson}
\{ q_i(x^1), p^k(y^1)\} =\delta^k_i \delta ,
\quad \delta = \delta (x^1 -y^1)\; .
\end{equation}
For the remaining variables $\bar q_i=(\omega_0,e^-_0,e^+_0)$
the canonical momenta $\bar{p_i}$ vanish (primary constraints).
The $\bar q$'s generate the secondary constraints
\begin{eqnarray}
G_1 &=& -p_2q_2+p_3q_3 +\partial_1p_1 \nonumber \\
G_2 &=& \partial_1p_2 +q_1p_2 - 
q_3 (V(p_1)+p_2p_3U(p_1)) \nonumber \\
G_3 &=& \partial_1 p_3-q_1p_3 + 
q_2 (V(p_1)+p_2p_3U(p_1)). \plabel{constraints}
\end{eqnarray}
Their Poisson brackets are \cite{kum97}
\begin{eqnarray}
\{ G_1,{G}_2\} &=& -G_2\delta \nonumber \\
\{ G_1,{G}_3\} &=& G_3 \delta \nonumber \\
\{ G_2,{G}_3\} &=& -[(V'(p_1)+(p_2)(p_3)U'(p_1))G_1+
\nonumber \\
 &\ & +(p_3)U(p_1)G_2+(p_2)U(p_1)G_3 ]\delta \plabel{algebra}\quad .
\end{eqnarray}

The constraints (\ref{constraints}) generate the gauge transformations
of the action (\ref{aacctt}):
\begin{eqnarray}
\delta z(p,q)&=&\{ z, G_i\} \xi^i ,  \\
\delta \bar q^k&=&-\dot \xi^k - \bar q^j{\cal C}_{ji}^k\xi^i
\plabel{g-trans1}
\end{eqnarray}
where $\xi^i$ is a parameter, $z(p,q)$ is an arbitrary function
of $p_i$ and $q^i$.
The structure functions ${\cal C}^k_{ij}$ are defined through
$\{ G_i,G_j\} ={\cal C}^k_{ij}G_k\delta$ by (\ref{algebra}).
There are no ternary constraints.

In the background field formalism one should decompose
all fields into quantum fluctuations and background values
\begin{equation}
e\to e+E,\quad \omega \to \omega +\Omega , \quad X\to X+Y \; ,
\plabel{back}
\end{equation}
where $E,\Omega ,Y$ denote background fields. Our gauge choice for the
fluctuations is
\begin{equation}
e_0^\pm =\omega_0 =0 \plabel{bgau}\; .
\end{equation}
Now the strategy is as follows. We subtract the classical action together
with all terms linear in the fluctuations. Next we integrate over quantum
fields. We anticipate the result that the effective action will be
just classical one. Hence it does not generate
any tadpole graphs. Otherwise, the following result would be true
at one loop only. From (\ref{back}) and (\ref{bgau})
 the action (\ref{aacctt}) is replaced by
\begin{eqnarray}
{\cal L}&\to&
e_1^-(-\partial_0-\Omega_0)X^+
+E_0^-\omega_1X^+ \nonumber \\
\ &\ &+e_1^+(-\partial_0+\Omega_0)X^-
-E_0^+\omega_1X^--\omega_1\partial_0X \plabel{ACT}\\
\ &\ &+(E_0^+E_1^--E_1^+E_0^-)V_2
+(E_0^+e_1^--E_1^+e_0^-)V_1 \; , \nonumber\\
V_1&=&V(X+Y)+U(X+Y)(X^++Y^+)(X^-+Y^-)-V(Y)-U(Y)Y^+Y^- \; , \nonumber \\
V_2&=&V_1-(V'(Y)+U'(Y)Y^+Y^-)X-U(Y)(X^+Y^-+X^-Y^+) \; . \nonumber
\end{eqnarray}
Integration over $\omega_1$, $e_1^-$ and $e_1^+$ gives
delta functions:
\begin{eqnarray}
\delta_{\omega_1}&=&\delta (-\partial_0 X+E_0^-X^+-E_0^+X^-) \nonumber \\
\delta_{e_1^-}&=&\delta (-\partial_0 X^+-\Omega_0X^++E_0^+V_1)
\nonumber \\
\delta_{e_1^+}&=&\delta (-\partial_0 X^-+\Omega_0X^--E_0^-V_1)
\plabel{delt}
\end{eqnarray}
For generic values of the background fields $E,\Omega$
the only solution with proper (vanishing) asymptotics is
\begin{equation}
X^\pm =X=0. \plabel{solu}
\end{equation}
Hence there are no quantum corrections except for two Jacobian factors.
The first one, which appears due to integration over $X$'s in the delta
functions becomes
\begin{equation}
\plabel{J1} J={\det}^{-1} (-\partial_0 \delta_i^k + M_i^k) \; , 
\end{equation}
\begin{equation}
M=
\begin{pmatrix} 0&E_0^-&-E_0^+ \\
E_0^+(V'(Y)+U'(Y)Y^+Y^-) & -\Omega_0+E_0^+U(Y)Y^- &E_0^+U(Y)Y^+ \\
-E_0^-(V'(Y)+U'(Y)Y^+Y^-)&-E_0^-U(Y)Y^-&\Omega_0-E_0^-U(Y)Y^+ 
\end{pmatrix} \; .
\end{equation}

The second Jacobian factor is the Faddeev--Popov determinant.

Taking (\ref{solu}) into account, one can easily find linearized
gauge transformations
\begin{eqnarray}
\delta \omega_1 &=&
-\dot{\xi^1}-\Epz(V'(Y)+Y^+Y^-U'(Y))\xi^2 +\Emz (V'(Y)+Y^+Y^-U'(Y))\xi^3 
\; , \nonumber \\
\delta \emo &=&
-\dot{\xi^2}-\Emz\xi^1 +\Omega_0 \xi^2-\Epz Y^-U(Y)\xi^2+\Emz Y^-U(Y)\xi^3
\; , \nonumber \\
\delta \epo &=&
-\dot{\xi^3}+\Epz\xi^1 -\Epz Y^+U(Y)\xi^2+\Emz Y^+U(Y)\xi^3-\Omega_0 \xi^3
 \nonumber \plabel{expl-delta}
 \quad .
\end{eqnarray}
This immediately gives us the Faddeev--Popov determinant 
\begin{equation}
J_{FP}=\det (\delta \bar q^i /\delta \xi^k) 
=\det (-\partial_0\delta_i^k -M_i^k ) \quad .
\plabel{FaPo}
\end{equation}
Notice that the arguments of the determinants in
(\ref{J1}) and (\ref{FaPo}) differ only by the
sign in front of the matrix $M_i^k$.
Introducing ghost fields $\bar c_i$, $c_k$ this determinant is generated by the
ghost action
\begin{equation}
\int ({\cal D}\bar c)({\cal D}c) 
\exp \left[i \int_M d^2x (-\bar c^i(-\partial_0\delta_i^k -M_i^k ) c_k)\right]
\plabel{act-gh}
\end{equation}
 
To evaluate the determinants (\ref{J1}) and (\ref{FaPo}) an IR
regularization is needed. According to our prescription
$\partial_0$ must be replaced by $\nabla_0$ in (\ref{J1}).
Consider 
\begin{equation}
-\ln J=\ln \det (-\nabla_0 \delta_i^k+M_i^k)=
\ln \det (-\nabla_0 )+\sum_{n=1} {\rm Tr}\left (
\nabla_0^{-1} M \right )^n \frac 1n\; .
\plabel{logdet}
\end{equation}
The first term in (\ref{logdet}) is independent of background
fields and will be neglected in what follows. The remaining terms are 
\begin{equation}
{\rm Tr} (\nabla_0^{-1}M)^n=
\int dx_1\dots dx_n {\rm tr} \left[ (\nabla_0^1)_{x_nx_1}M(x_1)
\dots (\nabla_0^1)_{x_{n-1}x_n}M(x_n)\right] .\plabel{Mn}
\end{equation}
Due to the presence of the step function $(x_n\le x_1\le \dots \le x_n)$
in $\nabla_0^{-1}$ only
coinciding points contribute to (\ref{Mn}). Hence we get
\begin{equation}
{\rm tr}\int dx \ln (1+\theta (0)M(x))\; .
\label{logM}
\end{equation}
The term (\ref{logM}) is not gauge invariant and thus would lead to
diffeomorphism and Lorentz anomalies. However 
we can use a freedom in choosing the IR regularization 
of the ghost action (\ref{act-gh}). The proper choice is to
replace $\partial_0$ by $\tilde\nabla_0$. With this regularization
we have
\begin{equation}
\ln J_{FP}=-\ln J
\end{equation}
and find that
all anomalous contributions to the effective action have cancelled.

\section*{Appendix B: Regularized Inverse Derivatives}
\label{AppB}
In the preceding sections we frequently encountered inverse derivative operators.
Here we shall define a proper infrared regularization scheme and list
the corresponding calculation rules which where used in the main text.
We restrict ourselves here to the case $x^0 \in {\rm R}$.
Two regularized Green functions $\nabla_{0xx'}^{-1}$ and 
$\tilde{\nabla}_{0xx'}^{-1}$ are introduced 
to replace $\partial_0^{-1}$ as 
\begin{equation}
\pz^{-1} \Rightarrow
\begin{cases}
 \lim_{\mu \to 0}\left(\pz -i\mu \right)^{-1}=
 \lim_{\mu \to 0}\left(\nabla_0^{-1} \right) \\
 \lim_{\mu \to 0}\left(\pz +i\mu \right)^{-1}=
 \lim_{\mu \to 0}\left(\tilde{\nabla}_0^{-1} \right) \quad
\end{cases}
\end{equation}
where $\mu =\mu_0 -i\varepsilon$. $\mu_0 \to +0$ 
represents the IR regularization, 
proper asymptotic behavior (cf. (\ref{inv}),(\ref{inv2}) below) is 
provided by $\varepsilon \to +0$.
 Note that a partial integration transforms
 $\nabla_0^{-1}$ into  $\tilde{\nabla}_0^{-1}$ and also that 
$\tilde{\nabla}_0^{-1}$ is not the complex conjugate of $\nabla_0^{-1}$.
The inverse operators are defined as the Green functions  $\nabla_0$ and
$\tilde{\nabla}_0$ and are calculated straightforwardly:
\begin{eqnarray}
\label{inv}
\left( \nabla_0^{-1} \right)_{xy}&=& -\theta (y-x)e^{i\mu (x-y)} 
\nonumber \\
\label{inv2}
\left(\tilde{\nabla}_0^{-1}\right)_{xy}&=& \theta (x-y)e^{-i\mu (x-y)} \quad 
\end{eqnarray}
$\theta$ denotes the step function.
The inverse squared operators are defined as the Green functions of
$(\nabla_0)^2$ and $(\tilde{\nabla}_0)^2$
and are given by
\begin{eqnarray}
\left( \nabla_0^{-2} \right)_{x,y}&=& (y-x)\theta (y-x)e^{i\mu (x-y)} 
\nonumber \\
\label{invsq}
\left(\tilde{\nabla}_0^{-2}\right)_{x,y}&=& (x-y)\theta (x-y)e^{-i\mu (x-y)}\quad .
\end{eqnarray}
Using (\ref{inv}) to (\ref{invsq}) the following identities may be verified 
easily:  
\begin{alignat}{2}
&\nabla_0^{-1} \nabla_0^{-1} =\nabla_0^{-2} & \qquad &
\tilde{\nabla}^{-1}_0 \tilde{\nabla}_0^{-1} =\tilde{\nabla}_0^{-2}  
\nonumber \\
&\nabla_0 \nabla_0^{-2} =\nabla_0^{-1} & \qquad &
\tilde{\nabla}_0 \tilde{\nabla}_0^{-2} =\tilde{\nabla}_0^{-1}  
\nonumber \\
&\nabla_0 \tilde{\nabla}_0^{-2}= \tilde{\nabla}_0^{-1} -2i\mu 
\tilde{\nabla}_0^{-2} & \qquad  &
\tilde{\nabla}_0 \nabla_0^{-2}= \nabla_0^{-1} +2i\mu \nabla_0^{-2}  \\
&\nabla_0 \tilde{\nabla}_0^{-1}=\delta(x-y) -2i\mu \tilde{\nabla}_0^{-1}  & 
\qquad &
\tilde{\nabla}_0 \nabla_0^{-1}=\delta(x-y) +2i\mu \nabla_0^{-1} 
\nonumber \\
&\nabla_{0xy}^{-2}= \nabla_{0(-x)(-y)}^{-2}= 
\tilde{\nabla}_{0yx}^{-2}  &\qquad &
\nabla_{0xy}^{-1}=-\tilde{\nabla}_{0yx}^{-1} \nonumber 
\end{alignat}

\noindent
Note that in the main text only these types of operations appeared 
and therefore
the limit $\mu \to 0$ does not cause any obviously divergent 
expressions, because 
$\mu$ does not appear with negative powers. Therefore this  
regularization scheme seems much superior to the one used in 
\cite{kum92}. Nevertheless, some care is necessary at $\mu \to 0$ 
in special cases (see below). 

This regularization introduces an IR-cutoff and thus eliminates
possible global quantum fluctuations which may occur in certain
backgrounds. For the classical solutions which enter our formulas to
zero order no such regularization  is necessary. Therefore in that
case a formal replacement
\begin{equation}
\int \nabla_{0xy}^{-1}\bar{F}(y^1)dx^0dx^1 \to x^0 \bar{F}(y^1)+
\bar{G}(y^1):=\tilde{x}^0
\plabel{eq132}
\end{equation}
is possible. Using the residual gauge transformations of the EF gauge
the complete r.h.s of (\ref{eq132}) may be even identified with a new
coordinate $\tilde{x}^0$. In quantum expressions we can expect 
(\ref{eq132}) to be true only for a restricted range of $x^0$ (or
$\tilde{x}^0$). A naive application of (\ref{inv}) e.g. in the
expression for $B_1^{(0)}$ ($\nabla_0 \bar{p}_{1,2}=0$)
\begin{equation}
B_1^{(0)}=\bar{p}_1+\nabla_0^{-1}\bar{p}_2
\plabel{eq133}
\end{equation}
for $\bar{p}_{1,2}=\hat{p}_{1,2}(x^1)e^{i\tilde{\mu}x^0}$
indeed yields an ill-defined (divergent) result even {\it before} the
regularization is removed. This can be circumvented by introducing a
different $\tilde{\mu}' \to \mu'-i\varepsilon'$ for homogenous
solutions as 
\begin{equation}
\bar{p}_{1,2}=\hat{p}_{1,2}(x^1)e^{i\tilde{\mu}'x^0} \; .
\end{equation}
Now (\ref{eq133}) is well-defined. Taking the limits in the sequence 
$\varepsilon' \to 0$, $\varepsilon \to 0$, $\mu \to 0$ for small $x^0$
then leads to the desired expression of type (\ref{eq132}) if a factor
$\frac{1}{\mu'}$ is absorbed in $\hat{p}_1$.

\vfill
\end{document}